\documentclass[fleqn,usenatbib]{mnras}

\usepackage{nccmath}
\usepackage{booktabs}
\usepackage{graphicx}	
\usepackage{amsmath,amssymb}	
\usepackage{makecell}

\newcommand{\logM}{\ensuremath{\mathrm{log}_{10} (M_{\star}/\mathrm{M}_{\odot})}}

\usepackage[T1]{fontenc}

\DeclareRobustCommand{\VAN}[3]{#2}
\let\VANthebibliography\thebibliography
\def\thebibliography{\DeclareRobustCommand{\VAN}[3]{##3}\VANthebibliography}

\usepackage{graphicx}	
\usepackage{amsmath}	
\usepackage{academicons}
\usepackage{threeparttable}
\usepackage{tablefootnote} 

\usepackage{orcidlink}

\title[Environmental quenching of low-mass galaxies]{JWST PRIMER: strong evidence for the environmental quenching of low-mass galaxies out to $\mathbf{\textit{z} \simeq 2}$ }
\author[M. L. Hamadouche]{M. L. Hamadouche\orcidlink{0000-0001-6763-5551}$^{1,2}$\thanks{E-mail: mhamadouche@umass.edu}, R. J. McLure$^{2}$, A. C. Carnall$^{2}$, D. J. McLeod\orcidlink{0000-0003-4368-3326}$^{2}$, J. S. Dunlop$^{2}$, 
\newauthor K. Whitaker$^{1,3}$, C. T. Donnan\orcidlink{0000-0002-7622-0208}$^2$, R. Begley\orcidlink{0000-0003-0629-8074}$^{2}$, T. M. Stanton\orcidlink{0000-0002-0827-9769}$^2$, O. Almaini$^{4}$, J. Aird$^{2}$, 
\newauthor F. Cullen${^2}$, S. Cutler\orcidlink{0000-0002-7031-2865}$^{1}$, A. M. Koekemoer\orcidlink{0000-0002-6610-2048}$^{5}$\\
$^{1}$Department of Astronomy, University of Massachusetts, Amherst, MA 01003, USA \\
$^{2}$SUPA\thanks{Scottish Universities Physics Alliance}, Institute for Astronomy, University of Edinburgh, Royal Observatory, Edinburgh, EH9 3HJ, UK \\
$^{3}$Cosmic Dawn Center (DAWN), Copenhagen, Denmark \\
$^{4}$School of Physics and Astronomy, University of Nottingham, University Park, Nottingham NG7 2RD, UK\\
$^{5}$Space Telescope Science Institute, 3700 San Martin Drive,
Baltimore, MD 21218, USA
}

\date{Accepted XXX. Received YYY; in original form ZZZ}

\pubyear{2015}

\begin{document}
\label{firstpage}
\pagerange{\pageref{firstpage}--\pageref{lastpage}}
\maketitle
\begin{abstract}
We present the results of a study investigating the galaxy stellar-mass function (GSMF), size-mass relations and morphological properties of star-forming and quiescent galaxies over the redshift range $0.25 < z < 2.25$, using the \textit{JWST} PRIMER survey. The depth of the PRIMER near-IR imaging allows us to confirm the double Schechter function shape of the quiescent GSMF out to $z \simeq 2.0$, via a clear detection of the upturn at $\logM\leq 10$ thought to be induced by environmental quenching. In addition to the GSMF, we confirm that quiescent galaxies can be split into separate populations at $\logM\simeq 10$, based on their size-mass relations and morphologies. We find that low-mass quiescent galaxies have more disk-like morphologies (based on S\'ersic index, Gini coefficient and $M_{20}$ metrics) and follow a shallower size-mass relation than their high-mass counterparts. Indeed, the slope of the size-mass relation followed by low-mass quiescent galaxies is indistinguishable from that followed by star-forming galaxies, albeit with a lower normalization. Moreover, within the errors, the evolution in the median size of low-mass quiescent galaxies is indistinguishable from that followed by star-forming galaxies (\hbox{$R_\mathrm{{e}} \propto (1+z)^{-0.25\pm 0.03})$}, and significantly less rapid than that displayed by high-mass quiescent galaxies (\hbox{$R_\mathrm{{e}} \propto (1+z)^{-1.14\pm 0.03})$}. Overall, our results are consistent with low and high-mass quiescent galaxies following different quenching pathways. The evolution of low-mass quiescent galaxies is qualitatively consistent with the expectations of external/environmental quenching (e.g. ram-pressure stripping). In contrast, the evolution of high-mass quiescent galaxies is consistent with internal/mass quenching (e.g. AGN feedback) followed by size growth driven by minor mergers.
\end{abstract}


\section{Introduction}
Understanding the different evolutionary pathways linking high-redshift galaxies with the bi-modal
population of star-forming and quiescent galaxies observed locally \citep[][]{Strateva2001,Baldry2004} remains a key goal in extragalactic astronomy. Addressing this question relies on being able to disentangle the influences of internal and external mechanisms responsible for the regulation of star-formation. 

Simulations have shown that feedback plays a crucial role in quenching star formation and producing the observed bi-modality of the local galaxy population \citep[e.g.][and references therein]{somerville_and_dave_quenching}. In the highest-mass halos ($\geq 10^{12}\ \mathrm{M_{\odot}}$) feedback from active galactic nuclei (AGN) is thought to play an important role in gas heating and the consequential quenching of star formation. In contrast, at early times, feedback in lower-mass halos is provided by supernovae, stellar winds and reionization \citep[see][]{dekelsilk1986,dekelandbirnboim2006}. At later times, environmental quenching mechanisms, such as ram-pressure stripping \citep{gun_gott1972}, and gas-rich mergers \citep[see e.g.][]{gabor_dave_morph_quenching,hopkins_mergers_simulations}, are thought to play an increasingly important role in quenching star formation in low-mass halos \citep{PengQuenching}.

From an observational perspective, the last two decades have seen considerable effort invested in improving our understanding of quiescent galaxies and quenching via studies of
the galaxy stellar-mass function (GSMF), the evolution of galaxy size and morphology with redshift and correlations between galaxy size, age and metallicity with stellar mass \citep[e.g.][]{muzzin_2013,davidzon_gsmf,vdw_sizes_candels,hamadouche22,hamadouche2023,beverage2023}. 

Based on the deepest available ground-based near-IR data, the wide-area study of \cite{mcleod2021_gsmf} showed that the quiescent GSMF at $z \leq 1.5$ follows a double Schechter function shape, with a characteristic upturn in the number density of $\logM\leq 10$ quiescent galaxies. Moreover, \cite{santini2022} 
found evidence for the same upturn at $z\simeq2-2.5$ in their combined CANDELS+Hubble Frontier Fields study. This upturn appears consistent with the emergence of a distinct population of environmentally-quenched galaxies \citep[e.g.][]{PengQuenching,papovich2018}, but whether or not this population of low-mass quenched galaxies emerged at still higher redshifts, remains an open question.

Previous studies of the GSMF have indicated that the rapidly evolving quiescent fraction means that while quiescent galaxies dominate the stellar-mass budget by $z \lesssim 0.5$, they 
account for only $\simeq 10\%$ of the mass budget at $z\simeq 3$. However, this result has been brought into question by the dramatically improved sensitivity and wavelength coverage provided by {\it JWST}. Recent {\it JWST-}based studies have spectroscopically confirmed high-mass quenched galaxies at $z\simeq5$ \citep[e.g.][]{carnall2024,degraaff2024} and have uncovered significantly higher number densities of high-mass quiescent galaxies at $z=3-5$ than suggested by pre-{\it JWST} studies \citep[e.g.][]{carnall2023,gould2023,valentino2023,antwidanso2023feniks,long2024}. Interestingly, at least some of these spectroscopically-confirmed, high-mass quiescent galaxies at $z\geq 3$ have stellar populations that can apparently only be explained by invoking much higher star-formation efficiencies than observed locally \citep[e.g.][]{carnall2024,glazebrook24}.

In addition to the GSMF, evidence for different quenching mechanisms should also be encoded in the evolution of galaxy size with redshift. In the local Universe, it is well established that quiescent galaxies are smaller than their star-forming counterparts (at fixed stellar mass) and follow steeper size-mass relations \citep{shen}. More recently, deep spectroscopic surveys have enabled size-mass relations to be determined for statistically representative samples of massive galaxies \citep{3dhst_vdw,beverage_metallicity_not_age,barone_2021_legac_SAMI}, building on earlier wide-field photometric studies at $z < 2$ \citep[e.g.][]{bruce12,bruce2014}. Within this context, in \citet{hamadouche22}, we selected robust samples of quiescent galaxies from the VANDELS \citep{vandels} and \hbox{LEGA-C} \citep{legac_survey_paper} spectroscopic surveys, finding that from $z = 1.3$ to $z = 0.6$, dry minor mergers are the dominant mechanism of size and stellar-mass growth at $\logM > 10.3$. 
The results of this spectroscopic study aligned extremely well with the results of a wide range of previous photometric studies of massive quiescent galaxies at $z < 2$ \citep[e.g.][]{trujillo2007,kriek2009,bruce12,bruce2014,cimatti_dry_merger,ross_size_2013,3dhst_vdw}.

However, investigations of the quiescent galaxy size-mass relation over a larger dynamical range of stellar mass paint a more 
complex picture; less-massive quiescent galaxies have sizes akin to star-forming galaxies at fixed stellar masses. As a result, many studies have suggested that the size-mass relation of quiescent galaxies is better described by a smoothly-broken or double power-law \citep[e.g.][]{lange2015,size_mass_nedkova,subaru_cam_size_mass,cutler2022}, with a pivot mass at $\logM \simeq 10$ \citep[e.g.][]{3dhst_vdw}.  
These observations highlight an important result; the flattening of the size-mass relation at lower stellar masses is an indication that environmental quenching mechanisms are becoming important, with consequences for the morphologies of low-mass quiescent galaxies \citep[e.g.][]{vandenbosch2008,kawinwanichakij2017}.

A galaxy's morphology, typically measured via the S\'ersic index ($n$), should also provide important information regarding its evolutionary history. Star-forming galaxies are known to have generally disk-like morphologies ($n\simeq 1$) that show little evolution with redshift \citep{patel2013}. In contrast, the S\'ersic indices of high-mass quiescent galaxies have been shown to evolve strongly with redshift \citep[e.g.][]{bruce12}. Moreover, massive post-starburst galaxies are known to be smaller than their quiescent counterparts, but exhibit similar S\'ersic indices, consistent with the expectations of rapid quenching \citep{almaini2017}. 

While much less is understood about the low-mass quiescent population, particularly at high redshift, \textit{JWST} has recently provided deep rest-frame optical and near-infrared imaging of these galaxies. Similar to trends observed at $z \sim 1$, the galaxy size-mass relation appears to flatten towards lower masses at $z > 1 $ \citep[e.g.][]{cutler2023,martorano24} and small numbers of low-mass quiescent galaxies have been discovered in over-dense regions at $z\simeq 2$ \citep[e.g.][]{sandles2023}. Together, these results suggest that environmental effects driving quenching at $\logM<10$ are not significantly changing the size or morphology of the quenching galaxies.

From the above discussion, it is clear that a powerful strategy for exploring the impact of different quenching mechanisms is to study both the GSMF and the evolution of galaxy size and morphology with redshift and stellar mass, for both star-forming and quiescent galaxies. Consequently, in this paper we investigate the GSMF for star-forming and quiescent galaxies, together with relationships between size, stellar mass and morphology, out to $z = 2.25$, using publicly available data from the \textit{JWST} PRIMER Survey \citep{primer}. The key motivation is to search for the signatures of environmental quenching within low-mass quiescent galaxies and to explore whether low and high-mass quiescent galaxies can be treated as distinct populations following different evolutionary pathways.

The structure of the paper is as follows. We first give details of the PRIMER dataset in Section \ref{datasectionc5}, and then discuss our fitting methods and sample selection in Section \ref{methodsssc5}. We then present our results on 
the GSMF and the relationships between size, stellar mass and morphology in Section \ref{resultsc5}, and discuss these findings in Section \ref{discussionc5}. Finally, we summarise our main conclusions in Section \ref{conclusionsc5}. 

Throughout this paper, we quote all magnitudes in the AB system and assume cosmological parameters of \hbox{$H_{0}$ = $\mathrm{70 \ {km} \ {s^{-1}} \ {Mpc^{-1}}}$}, \hbox{$\mathrm{\Omega_{m}} = 0.3$} and \hbox{$\mathrm{\Omega_{\Lambda}} = 0.7$}. We use a \cite{kroupaimf} initial mass function and the \cite{asplund2009} Solar abundance of \hbox{$\mathrm{Z}_{\odot}=0.0142$}.

\newpage

\begin{table*}
\setlength{\tabcolsep}{4.5pt}
\footnotesize
\renewcommand{\arraystretch}{1.0}
\caption[Details of parameter ranges \& priors for PRIMER]{Details of the parameter ranges and priors adopted for the {\scshape Bagpipes} fitting of the PRIMER photometry (see Section \ref{methodsssc5}). Priors listed as logarithmic are uniform in log-base-ten of the parameter.}
\label{tab:bagpipes_model}
    \begin{tabular}{lcccccc}
     \midrule
\textbf{Component} & \textbf{Parameter} & \makecell{\textbf{Symbol} / \\ \textbf{Unit}} & \textbf{Range} & \textbf{Prior} & \makecell{\textbf{Hyper-}\\ \textbf{parameters}}\\ 
  \midrule 
Global & Redshift & $z\mathrm{_{phot}}$ & \makecell{$z\mathrm{_{phot}}$ \\ \footnotesize $(\pm 0.015)$} &  \footnotesize Gaussian & \makecell{$\mu = z\mathrm{_{phot}}$ \\ $\sigma = 0.005$}\\
\midrule
SFH & \makecell{Stellar mass formed \\ Metallicity \\ Falling slope \\ Rising slope \\ Peak time} & \makecell{$M_{\star}/\mathrm{M_{\odot}}$ \\ $Z_{\star}/\mathrm{Z_{\odot}}$ \\ $\alpha$ \\  $\beta$ \\ $\tau $/ Gyr } & \makecell{(1,  10$^{13}$) \\(0.2, 2.5)  \\ (0.1, 10$^3$) \\ (0.1, 10$^3$)\\ (0.1,  $t\mathrm{_{obs}}$) } & \makecell{\footnotesize log\\ \footnotesize log \\ \footnotesize log \\ \footnotesize log \\ \footnotesize uniform} \\
\midrule
Dust & \makecell{5500 {\AA} attenuation\\ Deviation from Calzetti \\ 2175 {\AA}\ bump strength} & \makecell{$A_{V}/$mag \\ $\delta$  \\ \textit{B} } & \makecell{ \footnotesize (0, 4) \\ \footnotesize ($-0.3, 0.3$)\\ \footnotesize (0, 5)} & 
\makecell{\footnotesize uniform \\  \footnotesize Gaussian  \\ \footnotesize uniform}
& \makecell{ $\mu=0.0$ \\$\sigma = 0.1$\\ }\\
\midrule \\
\end{tabular}
\end{table*}

\begin{figure*}
    \centering
    \includegraphics[width = \linewidth]{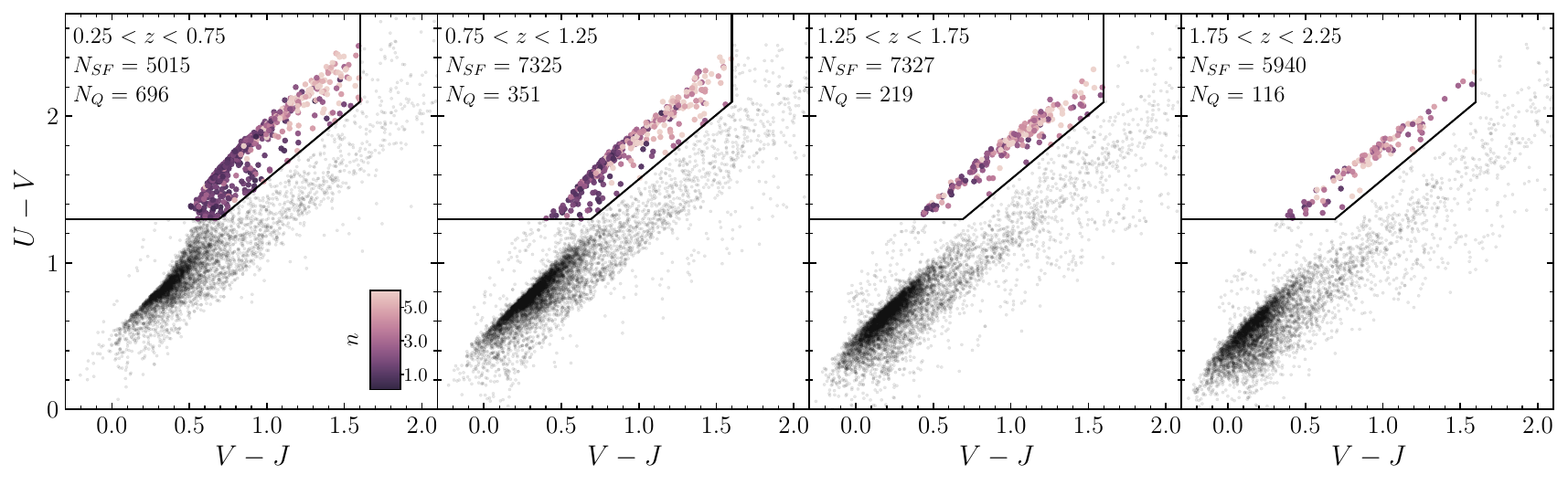}
    \caption[\textit{UVJ} diagrams for final PRIMER samples]{\textit{UVJ} diagrams as a function of redshift for the final, mass-complete, PRIMER UDS and COSMOS samples. In each panel, \textit{UVJ}-selected quiescent galaxies are colour-coded by S\'ersic index while star-forming galaxies are shown as black points. In the first three redshift bins, quiescent galaxies with redder colours are seen to have higher values of S\'ersic index. Small number statistics make it difficult to ascertain whether this trend persists in the highest-redshift bin.}
    \label{fig:primeruvj}
\end{figure*}

\section{Data}\label{datasectionc5}
In this section, we give a brief overview of the PRIMER survey data we make use of throughout this paper. 
\subsection{The PRIMER survey}
The Public Release IMaging for Extra-galactic Research (PRIMER) is a Cycle 1 Treasury Programme, providing deep \textit{JWST} NIRCam+MIRI imaging of the CANDELS COSMOS and UDS legacy fields. The NIRCam imaging is available in eight filters (F090W, F115W, F150W, F200W, F277W, F356W, F410M and F444W) covering the wavelength range \hbox{$0.9 - 5\ \mu$m}, with integration times varying between a minimum of 14 mins and a maximum of 84 mins. The PRIMER UDS and COSMOS fields also benefit from deep optical \textit{HST}/ACS imaging from the CANDELS survey \citep[][]{candels_grogin,candels_koekemoer} in the F435W, F606W and F814W filters. In this study, we makes use of the PRIMER area with full 
ACS+NIRCam coverage, which amounts to $\simeq 300$ sq. arcmin.

\subsection{Photometric catalogues}

For the PRIMER datasets, multi-wavelength catalogues are constructed using \textsc{SourceExtractor} \citep{bertin_arnouts_sextractor} in dual-image mode. To facilitate the selection of quiescent galaxies up to $z \sim 3$, F356W is used as the primary detection image. Isophotal photometry is performed on the PSF-homogenised images (PSF matched to F444W) and each of the catalogues requires a $5\sigma-$detection in the detection image, and a $3\sigma-$detection in at least one other band to minimise spurious detections. 

These catalogues were processed by various different photometric redshift codes and template sets, providing final median redshifts with a typical accuracy of $ \sigma_{\rm dz} \simeq 0.02$ and catastrophic outlier rates of only $\simeq3\%$ \citep[][]{Begley24}. We use these photometric redshifts during our \textsc{Bagpipes} fitting, discussed in the next section. 

\section{Methods and sample selection}\label{methodsssc5}

\subsection{Photometric fitting using \textsc{Bagpipes}}
We use {\scshape Bagpipes} \citep[][]{bagpipespaper} to fit the available photometric data for all galaxies in the initial PRIMER catalogues. A double-power-law star-formation history model is employed, incorporating the updated 2016 versions of the BC03 stellar population synthesis models \citep[][]{bruzualcharlot2003,charlot_chevallard_2016_BC03}, with stellar metallicity allowed to vary over the range $0.2 - 2.5\ \mathrm{Z}_{\odot}$, using a logarithmic prior. 

We use the \cite{Salim_2018} dust attenuation law, which parameterizes the dust curve through a power-law deviation, $\delta$, from the \cite{actualCalzettiLAW} attenuation law, as used in \citet{hamadouche2023}. Nebular continuum and emission lines are modelled using the {\scshape Cloudy} photo-ionization code \citep[][]{cloudycode}, using a method based on that of \cite{byler17}. We assume a fixed ionization parameter of $\mathrm{log_{10}}(U)= -3$. Full details of the free parameters and priors used in the fitting are provided in \hbox{Table \ref{tab:bagpipes_model}}.

\subsection{Selection of a mass-complete sample}\label{sampleselectprimerjades}
We first restrict our sample to those objects with photometry in $\geq 10$ bands, before imposing a 
$\chi^2 < 50$ cut on the SED fits to the photometry in order to remove artefacts, objects with corrupted photometry or spurious fits (e.g. fits which have hit the edges of the priors). We then use the rest-frame colours from \textsc{Bagpipes} to separate the star-forming and quiescent galaxies via the $UVJ$ diagram \citep{williams09bicolour}, requiring that quiescent galaxies have:
\begin{itemize}
    \setlength\itemsep{-0.1em}
    \item $U - V > 0.88 \times (V - J) + 0.69$
    \item $U - V > 1.3$
    \item $V - J < 1.6$
\end{itemize}

Additionally, to determine the effective stellar-mass limits of our PRIMER sample, we follow the procedure proposed in \cite{pozzetti2010}. For each galaxy, we calculate the limiting stellar mass that a galaxy would have if its apparent magnitude were equal to the limiting $5\sigma$ magnitude of the survey. Thus, the limiting stellar mass, $\mathrm{log}(\mathcal{M}_{\mathrm{lim}})$, of a single galaxy is given by:

\begin{equation}
    \mathrm{log}(\mathcal{M}_{\mathrm{lim}}) = \mathrm{log}(\mathcal{M}) + 0.4(m_\mathrm{F356W} - m_{\mathrm{lim}})
    \label{eqn:pozzetti}
\end{equation}

\noindent
where $\mathrm{log}(\mathcal{M})=\logM$ and $m_\mathrm{F356W} - m_{\mathrm{lim}}$ is the difference between the apparent F356W magnitude of the galaxy and the $5\sigma$ magnitude limit. 
We thus define the 90\% mass-completeness limit for the full sample at each redshift as the minimum stellar mass below which 90\% of the individual limiting stellar masses lie. 

The PRIMER COSMOS and UDS fields have variable depths across their shallow, medium and deep regions \citep{Donnan24}. 
After correcting to total, the median $5\sigma$~depth across the shallowest PRIMER regions is $\simeq$ 28.5 mag in the F356W selection filter. However, to ensure that objects in our sample have high enough signal-to-noise to provide robust size measurements, we adopt a more conservative limiting magnitude of $m_\mathrm{F356W} = 28$ for our final sample. The sample sizes and 90\% mass-completeness limits corresponding to this magnitude cut are presented in Table~\ref{tab:mass_completeness_numbers}.
\begin{table*}
\setlength{\tabcolsep}{4pt} 
\centering
\renewcommand{\arraystretch}{1.2}
\caption[Mass-completeness limits for PRIMER]{Sample sizes ($N$) and mass-completeness limits $(\mathcal{M}_{\mathrm{lim}})$ for the star-forming and quiescent galaxies in each $\Delta z=0.5$ redshift bin (see Section \ref{methodsssc5}). These sub-samples were used to derive our measurements of the galaxy stellar-mass functions (see Fig. \ref{fig:smfs}) and include galaxies that subseqeuntly failed to return robust size measurements.}
\label{tab:mass_completeness_numbers}
\begin{tabular}{l cc cc cc cc}
\toprule
    & \multicolumn{2}{c}{$0.25 < z < 0.75$}  & \multicolumn{2}{c}{$0.75 < z < 1.25$} & \multicolumn{2}{c}{$1.25 < z < 1.75$} &  \multicolumn{2}{c}{$1.75 < z < 2.25$} \\
\midrule

Galaxy type&  $N$ & $\mathrm{log}(\mathcal{M}_{\mathrm{lim}})$ & $N$ & $\mathrm{log}(\mathcal{M}_{\mathrm{lim}})$ & $N$ & $\mathrm{log}(\mathcal{M}_{\mathrm{lim}})$ & $N$ & $\mathrm{log}(\mathcal{M}_{\mathrm{lim}})$ \\
\midrule

Star forming & 5338 & 7.96 & 7855 & 8.11 & 7900 & 8.29 & 6358 & 8.46 \\
Quiescent & 1020 & 8.11 & 554 & 8.29 & 332 & 8.52 & 201 & 8.72 \\
\bottomrule
\end{tabular}
\end{table*}




\subsection{Size measurements and colour gradients}\label{chap5:sizegalfit}
We use {\scshape Galfit} \citep{galfit_paper} to measure the F356W-based sizes of the galaxies in our sample, adopting a similar procedure to \citet{hamadouche22}. As suggested by \cite{suess2022}, it is necessary to sample rest-frame near-infrared wavelengths in order to study the mass-weighted structural evolution of galaxies. The F356W imaging allows us to access the rest-frame near-infrared ($\lambda_\mathrm{{rest}} > 1.09 \ \mu\mathrm{m}$) over the full redshift range of our sample and is $\geq 0.5$ magnitudes deeper than the F410M and F444W imaging.

We set up an automated fitting routine (in \textsc{Python}) that uses the output from SourceExtractor \citep{bertin_arnouts_sextractor} as inputs for \textsc{Galfit}, in addition to generating the necessary image, segmentation and weight-map cutouts for each galaxy. The segmentation map is used to mask neighbouring galaxies from the fits, unless the neighbours are 
within $\pm 2.5$ magnitudes and separated by $\leq 3^{\prime\prime}$ from the target galaxy, in which case the target galaxy and neighbours are fit simultaneously.
We restrict the image-cutout size to 200 x 200 pixels ($6^{\prime\prime}\times 6^{\prime\prime}$ on the sky; $0.03^{\prime\prime}$/pix) and use an empirical point-spread function (PSF) created from a 
stack of isolated and unsaturated stars.

To avoid getting stuck in false local minima, we run \textsc{Galfit} over a grid of S\'ersic index and observed effective radius between $0.1 - 10$ and $0.025'' - 3''$, to incorporate the full range of realistic values. To ensure convergence, we implement a system during fitting to repeatedly `kick' \textsc{Galfit} out of local minima, before adopting the morphological parameters from the fit with the global minimum value of $\chi^2$. 

We also obtain non-parametric morphological measurements for our mass-complete samples using \textsc{StatMorph} \citep{rodriguezgomez2019}, which calculates parameters such CAS \citep[concentration, asymmetry and smoothness/clumpiness, see][for a review]{conselice_review}, the Gini coefficient $G$,  and the second-order moment of the brightest 20\% of a galaxy's pixels, $M_{20}$ \citep[see][]{lotz_2008}. 
\subsection{Final sample selection}
\begin{figure*}
     \centering
    \includegraphics[width = \linewidth]{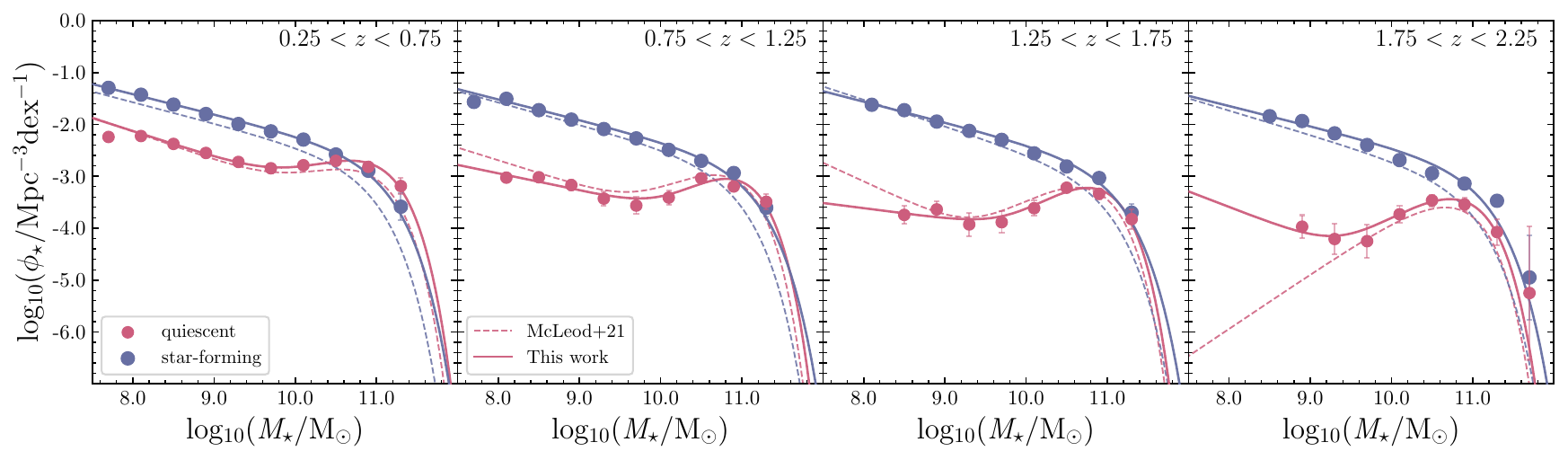}
    \caption[Star-forming \& quiescent galaxy number densities]{Galaxy stellar-mass functions for star-forming and quiescent galaxies within the PRIMER UDS and COSMOS fields over the redshift range $0.25 < z < 2.25$. The solid blue and red lines are our best-fitting single and double Schechter function fits to the star-forming and quiescent stellar-mass functions, respectively (see Table~\ref{tab:gsmf_fits}). The dashed lines are the best-fitting parameterizations from the wide-area, ground-based study of \protect\cite{mcleod2021_gsmf}. The depth and area of the PRIMER survey allows the low-mass upturn in the quiescent galaxy stellar-mass function to be clearly identified out to $z\geq 2$.}
    \label{fig:smfs}
\end{figure*}

We use the 90\% mass-complete sample presented in Table~\ref{tab:mass_completeness_numbers} to calculate the stellar-mass functions presented in Fig. \ref{fig:smfs}. However, to ensure a robust analysis of the sizes and morphologies of the PRIMER galaxies, we perform a final visual inspection, removing any galaxies for which the size fitting failed due to contamination by nearby stars/bright objects, or because the galaxy was too close to the image edge. This process removed $\simeq10$\% of the mass-complete sample, leaving a final, robust sample of $\simeq 1400$ quiescent and $\simeq 25000$ star-forming  galaxies over the redshift range $0.25 < z < 2.25$. The \textit{UVJ} diagrams for this sample as a function of redshift are presented in Fig.~\ref{fig:primeruvj}. 

\section{Results}\label{resultsc5}

In this section, we present the results obtained from our SED fitting and size measurements for our final PRIMER samples of star-forming and quiescent galaxies over the redshift range $0.25 < z< 2.25$.
\begin{table*}
\centering
\small
\renewcommand{\arraystretch}{1.0}
\caption{The best-fitting single (upper) and double (lower) Schechter function parameters to the observed star-forming and quiescent galaxy stellar-mass functions, where $\mathcal{M}_{\star} \equiv \logM$ and the units of $\phi_{\star_{1}}$ and $\phi_{\star_{2}}$ are dex$^{-1}$Mpc$^{-3}$. }
\label{tab:gsmf_fits}
\begin{tabular}{c|ccccc}
\toprule 
\multicolumn{6}{c}{star-forming} \\
\midrule
Redshift &  $\mathcal{M}_{\star}$ & $\phi_{\star_{1}}$ & $\alpha_{1}$ & $\phi_{\star_{2}}$ & $\alpha_{2}$ 
\\
\midrule 
$0.25 < z < 0.75$ &  $10.93 \pm 0.06$ & $-2.92 \pm 0.12$ & $-1.39 \pm 0.05$ & - &  - \\ 
$0.75 < z < 1.25$ & $10.95 \pm 0.04$ & $-3.05 \pm 0.09$  & $-1.40 \pm 0.04$ & - & - \\
$1.25 < z < 1.75$ & $10.94 \pm 0.05$ & $-3.11 \pm 0.11$ & $-1.40 \pm 0.05$ & - & -  \\
$1.75 < z < 2.25$ &  $10.97 \pm 0.03$  & $-3.17 \pm 0.09$ & $-1.39 \pm 0.06$& - & - \\
\midrule
\multicolumn{6}{c}{quiescent}\\
\midrule
Redshift &  $\mathcal{M}_{\star}$ & $\phi_{\star_{1}}$ & $\alpha_{1}$ & $\phi_{\star_{2}}$ & $\alpha_{2}$ 
\\
\midrule
$0.25 < z < 0.75$ & $10.70 \pm 0.16$ & $-2.70 \pm 0.10$ & $0.30 \pm 0.43$ & $-3.46 \pm 0.34$ & $-1.30 \pm 0.12$ \\
$0.75 < z < 1.25$ & $10.73 \pm 0.14$ & $-3.00 \pm 0.06$ & $0.19 \pm 0.45$ &$-4.55\pm 0.45$ & $-1.53 \pm 0.20$ \\
$1.25 < z < 1.75$ & $10.67 \pm 0.13$ & $-3.17 \pm 0.06$ & $0.22 \pm 0.46 $ & $-5.07 \pm 1.09$ & $-1.54 \pm 0.56$\\
$1.75 < z < 2.25$ &  $10.71 \pm 0.14$ & $-3.37 \pm 0.09$ & $0.02 \pm 0.38$ & $-7.25 \pm 4.20$ & $-2.60 \pm 2.25$ \\
\bottomrule
\end{tabular}
\end{table*}

\subsection{The galaxy stellar-mass function}\label{gsmfc5}
In this analysis, we use the depth and wavelength coverage of the PRIMER imaging to improve on previous
studies of the GSMF based on ground-based and \textit{HST} imaging. We derive robust stellar masses from \textsc{Bagpipes} and calculate number densities for our PRIMER samples of star-forming and quiescent galaxies between $z = 0.25$ and $z = 2.25$, in bins of width $\Delta z=0.5$.

Following \cite{mcleod2021_gsmf}, we fit the observed star-forming and quiescent GSMFs with single and double-Schechter functions, respectively. The single-Schechter function takes the form:

\begin{equation}\label{singleschechter}
    \phi(M) = \phi_{\star} \cdot \mathrm{ln(10)} \cdot 10(M - M_{\star})^{(1 + \alpha)} \cdot \mathrm{exp}[-10(M - M_{\star})] 
\end{equation}

\noindent and the double-Schechter function:

\begin{align}\label{doubleschechter}
\phi(M)  = & \ \mathrm{ln(10)} \cdot \mathrm{exp}[-10(M - M_{\star})] \nonumber  \\
     & \cdot 10^{(M - M_{\star})} \cdot [ \phi_{\star_{1}} \cdot 10^{(M - M_{\star})\alpha_{1}}\nonumber \\
     & + \ \phi_{\star_{2}}  \cdot 10^{(M - M_{\star})\alpha_{2}}]
\end{align}

\noindent
where in both functions, the characteristic stellar mass is given by $M_{\star}$.

We plot our new determinations of the quiescent and star-forming GSMFs, along with the best-fitting Schechter functions, in Fig. \ref{fig:smfs}. The corresponding best-fitting Schechter function parameters are reported in Table \ref{tab:gsmf_fits}.
For comparison, we also plot the best-fitting GSMFs derived in \cite{mcleod2021_gsmf}, based on 3 sq. degrees of the deepest available ground-based near-IR imaging. 

It can be seen from Fig. \ref{fig:smfs} that, overall, our new GSMF results are in good agreement with those of \cite{mcleod2021_gsmf}. However, the extra depth provided by the PRIMER imaging allows us to confirm that the quiescent GSMF follows a double Schechter function shape over the full redshift range of our sample. The PRIMER NIRCam imaging employed in this study is $\simeq 3$ magnitudes deeper than the deepest available ground-based near-IR imaging. Consequently, while \cite{mcleod2021_gsmf} found no strong evidence for a low-mass upturn in the quiescent GSMF at $z\geq 1.5$, we have firmly established that the upturn is present out to $z\simeq 2.25$, in excellent agreement with the results presented in \cite{santini2022}. 

The low-mass upturn in the quiescent galaxy stellar-mass function is widely interpreted as due to the emergence of a separate population of low-mass, environmentally-quenched galaxies \citep{PengQuenching}. Within this context, our GSMF results immediately suggest that the quiescent galaxy population can be split into two separate populations at $\logM\simeq 10$ and that the low-mass quiescent galaxy population was becoming established as early as $z\simeq 2$.

\subsection{Galaxy size-mass relations at \textbf{0.25 $<$ \textit{z} $<$ 2.25}} \label{PJsizemassrelations}
Adopting the same redshift bins employed for the GSMF, we fit size-mass relations for our mass-complete samples of star-forming and quiescent galaxies. Following \cite{hamadouche22}, we fit a single power-law relation as a function of galaxy mass, such that:

\begin{equation}
    \mathrm{log}_{10}\bigg(\dfrac{R_e}{\mathrm{kpc}}\bigg) = \alpha \times \mathrm{log}_{10}\bigg(\dfrac{M_\star}{5 \times 10^{10} \mathrm{M}_{\odot}}\bigg) + \mathrm{log}_{10}(A)
    \label{sizemass_linear}
\end{equation}

\noindent
where $R_{\mathrm{e}}$ is the effective radius of the galaxy, $\alpha$ is the slope and $\mathrm{log}_{10}(A)$ is the normalisation of the relationship. We fit the single power-law relation to the full star-forming galaxy sample and separately to low-mass and high-mass sub-samples of the quiescent galaxies split at $\logM=10$, as motivated by the observed inflection point in the quiescent GSMF. In addition, we fit a smoothly-broken power law over the entire stellar-mass range for quiescent galaxies, defined as:

\begin{equation}
     R_e =  R_p \bigg(\dfrac{M_\star}{\mathrm{M}_{p}}\bigg)^{\alpha} + \Bigg[\dfrac{1}{2} \bigg\{1 + \bigg(\dfrac{M_\star}{\mathrm{M}_{p}}\bigg)^{\delta} \bigg\}\Bigg]^{ \beta - \alpha/ \delta}
     \label{eqn:sizedbplplaw}
\end{equation}

\noindent
where $\alpha$ and $\beta$ are the low and high-mass power-law slopes, ${\rm M}_{p}$ is the pivot mass and $R_{p}$ is the corresponding size for a given pivot mass. 

When fitting the size-mass relations, we choose a constant uncertainty on galaxy sizes of $0.1\,$dex, due to the uncertainties provided by \textsc{Galfit} being significantly underestimated \citep{haussler_galfit_errs}. The F356W images provide physical resolution almost the same as the \textit{HST} F160W images used to measure sizes in \cite{hamadouche22}. In this earlier work, the size measurements for VANDELS quiescent galaxies agreed with those from 
\cite{vdw_sizes_candels,3dhst_vdw} to within $\pm0.1$ dex. The constant error adopted for the fitting presented in this paper therefore provides a more realistic estimate of the typical uncertainty in galaxy sizes \citep[e.g.][]{ross_size_2013,hamadouche22}. 

Overall, we find that our rest-frame near-IR sizes are $\simeq 0.15$~dex smaller than $z\leq3$ sizes measured in the rest-frame optical, consistent with results reported by numerous previous studies \citep[e.g.][]{mowla_cosmos_dash,suess2022,kartaltepe2023,cutler2023,martorano24}.

\begin{figure*}
    \includegraphics[width = \linewidth]{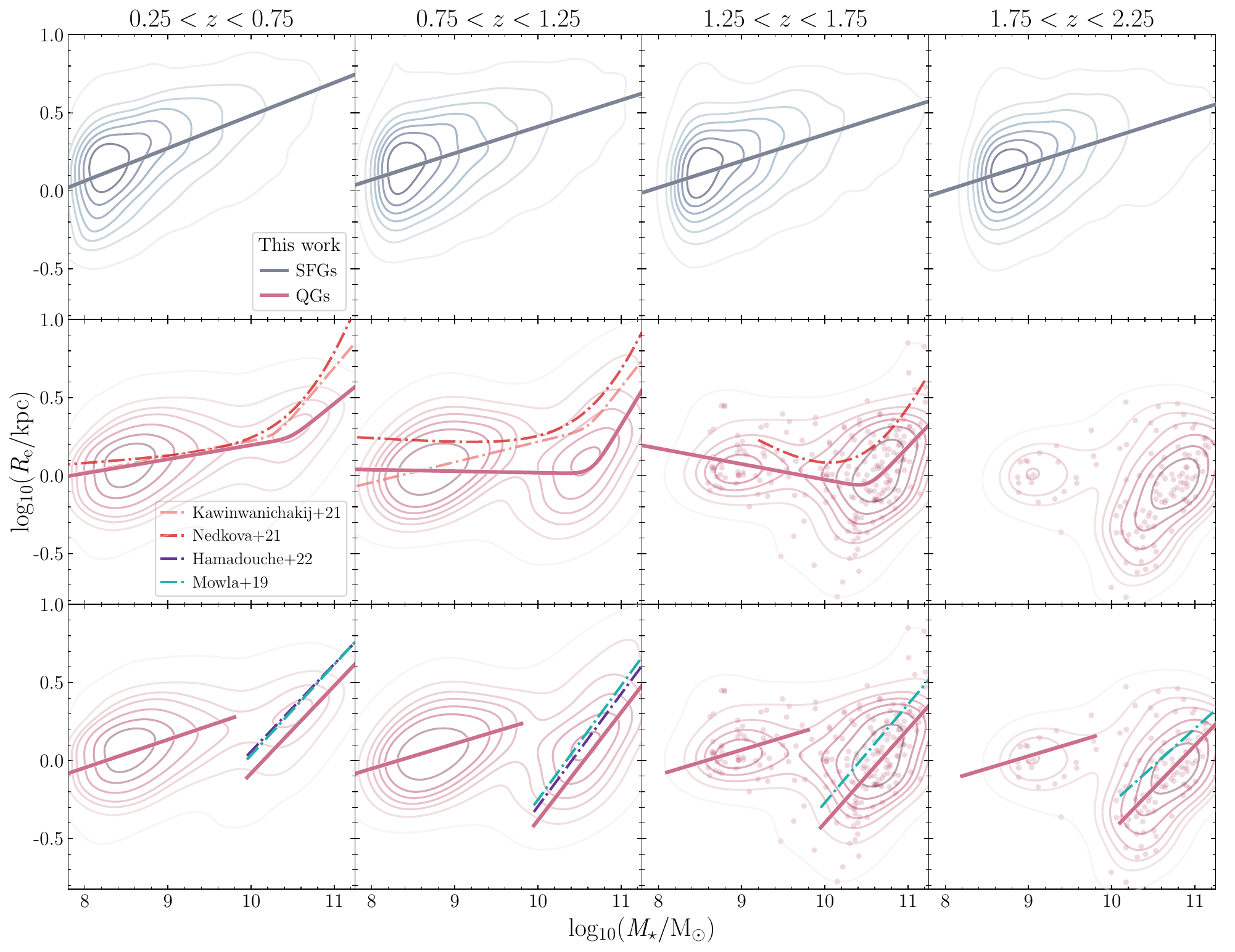}
    \caption{Star-forming (blue, top panel) and quiescent (red, middle and bottom panels) galaxy size-mass relations. Blue and red contours indicate the sizes of PRIMER galaxies at $0.25 < z < 2.25$, measured in the F356W filter. In the middle panel, we compare our double power-law fits to previous literature results \protect\citep{subaru_cam_size_mass,size_mass_nedkova}. It can be seen that the double power-law fits do not capture the full mass evolution across our redshift range. In the bottom panel we show our single power-law fits to the low-mass and high-mass quiescent galaxy sub-samples and compare to previous literature fits to high-mass quiescent galaxies \protect\citep{mowla_cosmos_dash,hamadouche22}. Our slopes for the high-mass end of the quiescent galaxy distribution are in excellent agreement with previous literature results, albeit with lower normalisations consistent with rest-frame near-IR sizes being smaller than rest-frame optical sizes.}
    \label{fig:size_mass_primer}
\end{figure*}

\subsubsection{Star-forming galaxies}
In the top panel of Fig. \ref{fig:size_mass_primer}, we present the size-mass distributions and relations for star-forming galaxies from $z = 0.25$ to $z = 2.25$. From Fig. \ref{fig:size_mass_primer}, and the results presented in Table \ref{tab:sizemassrelations}, it is clear that the normalisation of the star-forming size-mass relation decreases with increasing redshift, consistent with galaxies having larger sizes at later times. Additionally, our results suggest that there is no strong evolution in the slope of the star-forming galaxy size-mass relation from $z = 2.25$ to $z = 0.25$, consistent with previous literature results that also fit the size-mass relation with a shallow slope of $\alpha \simeq 0.2$ \citep[e.g.][]{3dhst_vdw,mowla_cosmos_dash}. The normalisation of the star-forming galaxy size-mass relation increases by only $\sim 0.15$ dex between $z = 2.25$ and $z = 0.25$, compared to the $\sim 0.40$ dex increase observed for high-mass quiescent galaxies. 

\subsubsection{Quiescent galaxies}
In the second row of Fig. \ref{fig:size_mass_primer}, we present our double power-law fits to the size-mass relations for the entire quiescent galaxy distribution. These fits could only be performed at $z <1.75$, due to small number statistics in the highest-redshift bin. A clear flattening in the low-mass slope of the quiescent galaxy size-mass relation can be seen, consistent with the results of other studies over the same redshift range \citep[e.g.][]{3dhst_vdw,size_mass_nedkova,subaru_cam_size_mass,cutler2022,martorano24}. 

Upon inspection, we find that our double power-law fits do not accurately describe the quiescent galaxy population as a whole. Although the high-mass slopes determined by \cite{size_mass_nedkova} and \cite{subaru_cam_size_mass} broadly agree with our fits, they do not account for the galaxies in the low-mass end of the high-mass sample, or the apparent stellar-mass gap between the low-mass and high-mass quiescent galaxies. 

We therefore perform separate single power-law fits on the low-mass and high-mass quiescent galaxy samples, to capture their true underlying distribution. If low-mass and high-mass quiescent galaxies are indeed quenching via different mechanisms, and thus proceeding along different evolutionary pathways, it is entirely reasonable to treat these sub-populations separately. The results of our single power-law fits to the quiescent galaxies are shown in the bottom panel of \hbox{Fig. \ref{fig:size_mass_primer}}, with the best-fitting parameters listed in Table~\ref{tab:sizemassrelations}. 

\begin{table*}
\setlength{\tabcolsep}{4pt} 
\centering
\small
\renewcommand{\arraystretch}{1.0}
\caption[Best-fitting parameters for single-power law fits]{The best-fitting parameters for the single power-law fits to the galaxy size-mass distributions, as described in Section \ref{PJsizemassrelations} and shown in the top and bottom panels of Fig \ref{fig:size_mass_primer}.}
\label{tab:sizemassrelations}
\begin{tabular}{c|cc|cc|cc}
\toprule
~ & \multicolumn{2}{c}{star-forming}  &  \multicolumn{2}{c}{low-mass quiescent}  & \multicolumn{2}{c}{high-mass quiescent} \\ 
\midrule
Redshift & $\alpha$ & $\mathrm{log}(A)$ & $\alpha$ & $\mathrm{log}(A)$ & $\alpha$ & $\mathrm{log}(A)$ \\
\midrule
$0.25 < z < 0.75$ & $0.21 \pm 0.01$ & $0.63 \pm 0.01$ & $0.18 \pm 0.02$ & $0.44 \pm 0.05$ & $0.56 \pm 0.08$ & $0.31\pm 0.05$  \\
$0.75 < z < 1.25$ & $0.17 \pm 0.01$ & $0.53 \pm 0.01$ & $0.17 \pm 0.02$ & $0.42 \pm 0.03$ & $0.69 \pm 0.08$ & $0.10 \pm 0.02$  \\
$1.25 < z < 1.75$ & $0.17 \pm 0.01$ & $0.48 \pm 0.02$ & $0.17^{a}$ & $0.34 \pm 0.02$ & $0.60 \pm 0.07$ & $0.02 \pm 0.02$ \\
$1.75 < z < 2.25$ & $0.17 \pm 0.01$ & $0.46 \pm 0.02$ & $0.17^{a}$ & $0.30 \pm 0.04$ & $0.55 \pm 0.08$ &  $ -0.07 \pm 0.02$  \\
\bottomrule
\end{tabular} \\
\vspace{2px}
\small $^a$For the two highest-redshift bins, the low-mass quiescent slopes were fixed to their low-redshift values.

\end{table*}

\begin{table*}
\setlength{\tabcolsep}{4
pt} 
\centering
\small
\renewcommand{\arraystretch}{1.0}
\caption[Best-fitting parameters for double-power law fits]{The best-fitting parameters for the double power-law fits to the quiescent galaxy size-mass distributions. Following \protect\cite{mowla2019a} and \protect\cite{subaru_cam_size_mass}, to minimise the degeneracy between $\delta$ (which controls the sharpness of the transition between the high and low-mass slopes) and the slopes, we use a constant value of $\delta = 6$. We do not fit the double power-law in the highest-redshift bin due to the low number of high-mass quiescent galaxies at $z \geq 1.75$.}
\label{tab:dblplawtable}
\begin{tabular}{c|ccccc}
\toprule
Redshift & $r_{\mathrm{p}}$ & $\alpha$ & $\beta$ & $M_{\mathrm{p}}$ 
\\
\midrule
$0.25 < z < 0.75$ & $1.80 \pm 0.16$ & $0.09 \pm 0.01$ & $0.43 \pm 0.14$ & $10.48 \pm 0.18$ \\
$0.75 < z < 1.25$ & $1.14 \pm 0.08$ & $-0.01 \pm 0.02$ & $0.89 \pm 0.23$ & $10.65 \pm 0.10$ \\
$1.25 < z < 1.75$ & $0.90\pm 0.06$ &  $-0.10 \pm 0.03$ & $0.56 \pm 0.12$ & $10.53 \pm 0.10$ \\
\bottomrule
\end{tabular}
\end{table*}

The bottom panel of Fig. \ref{fig:size_mass_primer} suggests that the quiescent galaxy population is indeed separated into two sub-populations, split at $\logM\simeq10$, fully consistent with the GSMF results presented in Section \ref{gsmfc5} 
and potentially signifying two distinct evolutionary channels for 
low-mass and high-mass quiescent galaxies (i.e. environmental vs. mass quenching).

In all four redshift bins, it is notable that the slope of the low-mass quiescent galaxy size-mass relation is shallower than the higher-mass relation. The low-mass slope mirrors that of the star-forming galaxies, albeit with a lower normalisation, and is consistent with quiescent galaxies displaying smaller sizes than star-forming galaxies at fixed stellar mass. This may point to the progenitors of these low-mass quiescent galaxies being star-forming galaxies that are smaller on average than the overall star-forming population. 

In contrast, the slope of the size-mass relation for higher-mass quiescent galaxies is steeper than that of lower-mass quiescent and star-forming galaxies. Moreover, we note that the high-mass slope of the double power-law and single power-law fits agree within $1 \sigma$ ($\alpha \simeq 0.5$), and are consistent with the results of several recent studies of the size-mass relation in massive quiescent galaxies at $z >1$ \citep[e.g.][]{3dhst_vdw,mowla_cosmos_dash,hamadouche22}. 
These findings indicate that star-forming, low-mass and high-mass quiescent galaxies occupy distinct regions of the size-mass plane, consistent with recent results at cosmic noon from \cite{cutler2023}. In Section \ref{envquench?}, we investigate possible quenching mechanisms which could reproduce the observed trends.

\subsection{Galaxy morphology}

\subsubsection{S\'ersic index}

\begin{figure*}
    \centering
    \includegraphics[width = \linewidth]{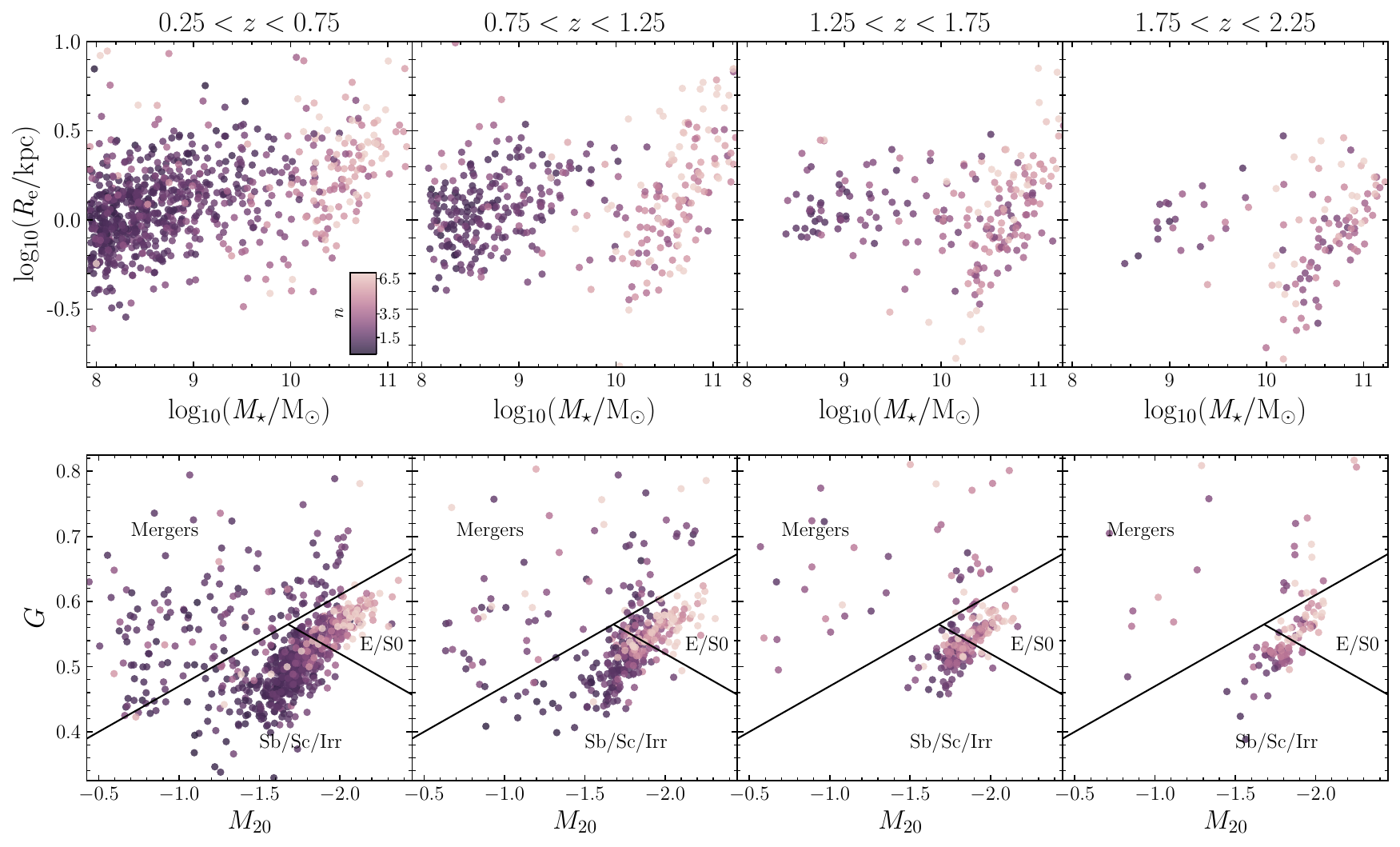}
    \caption{The top panel shows our quiescent galaxy sample on the size-mass plane, colour-coded by S\'ersic index. The bottom panel shows the same sample on the Gini-$M_{20}$ plane, again color-coded by S\'ersic index. The distribution of the sample on the $G-M_{20}$ plane provides a non-parametric method of classifying the galaxies as ellipticals (E0), spirals (Sb, Sc), irregulars (Irr) or mergers \citep{lotz_2008}. On a galaxy-by-galaxy basis, high-mass quiescent galaxies exhibit higher S\'ersic indices than low-mass quiescent galaxies up to $z \simeq 2.25$. Higher-mass, higher-S\'ersic index quiescent galaxies appear to lie mostly in the E0/S0 region of the $G-M_{20}$ plane, whereas low-mass, low-S\'ersic index quiescent galaxies mostly lie in the Sb/Sc region.}
    \label{fig:sizemasssersic}
\end{figure*}

In the top panel of Fig. \ref{fig:sizemasssersic} we show the size-mass distribution of PRIMER quiescent galaxies colour-coded by S\'ersic index and present the median values of size, S\'ersic index and axis ratio as a function of redshift in Table~\ref{tab:sizesersic}. In our sample, the low-mass quiescent galaxies clearly have much lower S\'ersic indices than their high-mass counterparts, and this trend is observable up to $z\leq 2.25$. It is unlikely that this clear difference is purely a result of increasing stellar mass, considering that high-mass star-forming galaxies do not have such extreme S\'ersic indices. This result confirms for the first time, over a wide redshift range, that low-mass and high-mass quiescent galaxies are indeed morphologically different populations. 

It can be seen from Table \ref{tab:sizesersic} that massive star-forming galaxies do have higher S\'ersic indices than the overall star-forming population, which may be indicative of the formation of a more prominent bulge component in comparison to lower-mass star-forming galaxies. However, this difference in S\'ersic index with mass is much less dramatic than the observed distinction between the low-mass and high-mass quiescent galaxies. Indeed, Table \ref{tab:sizesersic} also indicates that low-mass quiescent galaxies have higher S\'ersic indices than star-forming galaxies of the same mass. This result may suggest that the S\'ersic indices displayed by the lower-mass quiescent galaxies are associated with morphologies transitioning from disc-like to lenticular/S0 galaxies. In Section \ref{discussionc5}, we discuss these results in the context of quenching and galaxy evolution.

\subsubsection{Non-parametric statistics}

We use \textsc{Statmorph} to calculate non-parametric morphological statistics, including the Gini coefficient ($G$) and $M_{20}$, the second-order moment of the brightest 20\% of pixels. 
In the bottom panel of Fig. \ref{fig:sizemasssersic}, we show the distribution of our quiescent galaxies on the $G-M_{20}$ plane, colour-coded by S\'ersic index, with the separating regions as defined in \cite{lotz_2008}. Galaxies with higher S\'ersic indices appear to mostly populate the E/S0 region of the $G-M_{20}$ plane, whilst galaxies with lower S\'ersic indices, and lower stellar masses, mostly populate the Sb/Sc/Irr region. Although not shown on the diagram, the star-forming galaxies in our sample mainly occupy the spiral region, suggesting that low-mass quiescent galaxies possess morphologies more similar to the star-forming population than the high-mass quiescent galaxy population. 
\begin{table*}
\setlength{\tabcolsep}{3pt} 
\small
\renewcommand{\arraystretch}{1.0}
\caption[Median sersic and sizes]{Median S\'ersic index, axial ratio and effective radius for the star-forming and quiescent galaxy samples, within each $\Delta z=0.5$ bin. We report figures for the full sample, together with the low and high-mass sub-samples.}

\label{tab:sizesersic}
\begin{tabular}{c|ccc|ccc|ccc}
\toprule
\textbf{star-forming} & \multicolumn{3}{c}{\textbf{all}} & \multicolumn{3}{c}{$\logM < 10.0$} & \multicolumn{3}{c}{$\logM \geq 10.0$}  \\ 
\midrule
Redshift & $n$ & $q$ & $R\mathrm{_{e}/kpc}$ & $n$ & $q$ & $R\mathrm{_{e}/kpc}$ & $n$ & $q$ & $R\mathrm{_{e}/kpc}$  \\
\midrule
$0.25 < z < 0.75$ & $1.20 \pm 0.02$ & $0.50\pm0.01$ & $1.72 \pm 0.06$ & $1.17\pm 0.01$ & $0.48 \pm 0.01$ & $1.68\pm 0.08$ & $1.93 \pm 0.13$ & $0.53 \pm 0.02$ & $3.37 \pm 0.35$ \\

$0.75 < z < 1.25$ & $1.26 \pm 0.01$ & $0.48 \pm 0.01$ & $1.59 \pm 0.08$ & $1.22 \pm 0.01$ & $0.47 \pm 0.01$ & $1.54 \pm 0.80$ & $1.91 \pm 0.09$ & $0.52 \pm 0.01$ & $3.01 \pm 0.27$  \\

$1.25 < z < 1.75$ & $1.33 \pm 0.01$ & $0.45 \pm 0.01$ & $1.50 \pm 0.10$ & $1.30 \pm 0.03$ & $0.45 \pm 0.01$ & $1.44 \pm 0.10$ & $1.94 \pm 0.08$ & $0.55 \pm 0.01$ & $2.69 \pm 0.20$  \\

$1.75 < z < 2.25$ & $1.34 \pm 0.02$ & $0.44 \pm 0.01$ & $1.47 \pm 0.08$ & $1.32 \pm 0.02$ &$0.43 \pm 0.01$ & $1.42 \pm 0.10$ & $1.68 \pm 0.08$ & $0.53 \pm 0.01$ & $2.45 \pm 0.44$  \\
\midrule
\textbf{quiescent} &  \multicolumn{3}{c}{\textbf{all}}  & \multicolumn{3}{c}{$\logM< 10.0$} & \multicolumn{3}{c}{$\logM \geq 10.0$} \\
\midrule
Redshift  & $n$ & $q$ & $R\mathrm{_{e}/kpc}$ & $n$ & $q$ & $R\mathrm{_{e}/kpc}$ & $n$ & $q$ & $R\mathrm{_{e}/kpc}$  \\
\midrule
$0.25 < z < 0.75$ & $1.62 \pm 0.07$ & $0.64 \pm 0.01$ & $1.30 \pm 0.07$ & $1.44 \pm 0.05$ & $0.65 \pm 0.01$ & $1.19\pm 0.08$ & $5.25 \pm 0.20$ & $0.61 \pm 0.02$ & $1.95 \pm 0.40$ \\

$0.75 < z < 1.25$ & $2.38 \pm 0.12$ & $0.63 \pm 0.01$ & $1.19 \pm 0.50$ & $1.66 \pm 0.11$ & $0.63 \pm 0.01$ & $1.17 \pm 0.78$ & $5.15\pm 0.20$ & $0.65 \pm 0.02$ & $1.44 \pm 1.00$ \\

$1.25 < z < 1.75$ & $3.53 \pm 0.03$ & $0.68 \pm 0.01$ & $1.11 \pm 0.07$ & $2.04 \pm 0.23$ & $0.71 \pm 0.03$ & $1.12 \pm 0.25$ & $4.37 \pm 0.20$ & $0.67 \pm0.02$ & $1.08 \pm 0.10$ \\

$1.75 < z < 2.25$ & $3.90 \pm 0.18$ & $0.68 \pm 0.02$ & $0.90 \pm 0.14$ & $2.36 \pm 0.31$ & $0.72 \pm 0.04$ & $0.97 \pm 0.10$ & $4.22 \pm 0.25$ & $0.66 \pm 0.03$ & $0.89 \pm 0.08$ \\
\bottomrule
\end{tabular}
\end{table*}

\begin{figure*}
    \centering
    \includegraphics[width=\linewidth]{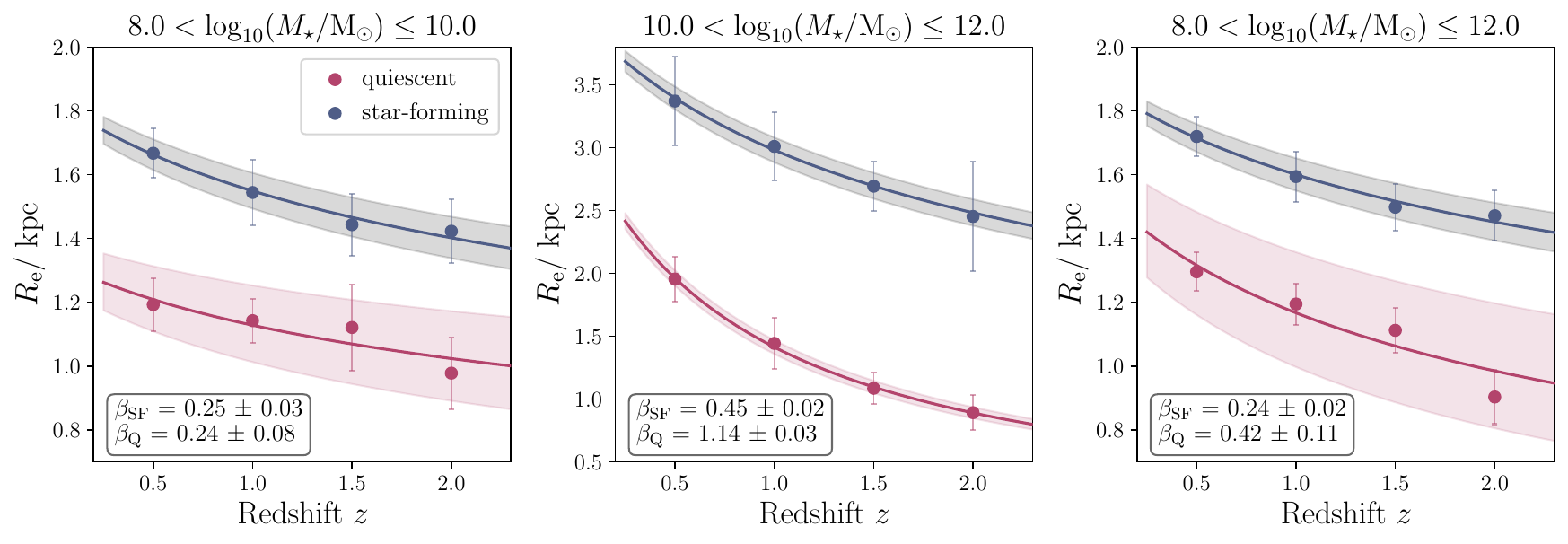}

    \caption{The size evolution of quiescent and star-forming galaxies within three different stellar-mass ranges. We find that the slope of the relation followed by low-mass quiescent galaxies is indistinguishable from that followed by low-mass star-forming galaxies. We also find that high-mass star-forming galaxies evolve more rapidly than low-mass star-forming galaxies. However, at high stellar masses it is clear that quiescent galaxies experience much more rapid size evolution that their star-forming counterparts.}
    \label{fig:size_redshift_evol}
\end{figure*}

\subsubsection{The evolution of median galaxy size}\label{sizeevolsec}
The size evolution of galaxies can be parameterised by a power-law relation of the form:
\begin{equation}\label{evol_eq}
    R_{e} = B_z \times (1 + z)^{-\beta}
\end{equation}

We fit this relation to the median sizes within each redshift bin for the star-forming and quiescent galaxies over the full stellar-mass range,  and separately between $8 < \logM < 10$ and $10 < \logM < 12$. Overall, our best-fitting relations (see Table~ \ref{tab:size_evol}) are in very good agreement with previous literature results \citep{subaru_cam_size_mass, cutler2022,3dhst_vdw}. As shown in Fig. \ref{fig:size_redshift_evol}, we find that the sizes of low-mass quiescent galaxies evolve as $R_\mathrm{{e}} \propto (1+z)^{-0.24\pm 0.08}$, which is the same rate as the evolution of star-forming galaxies in the same stellar-mass range; $R_\mathrm{{e}} \propto (1+z)^{-0.25\pm 0.03}$. However, the evolution of low-mass quiescent galaxies is much slower than the $R_\mathrm{{e}} \propto(1+z)^{-1.14\pm 0.02}$ followed by high-mass quiescent galaxies. These results suggest that the population of low-mass quiescent galaxies are drawn from a parent population of low-mass (dwarf) star-forming galaxies. 

Based on a sample of $0.5 < z < 2.5$ galaxies with \textit{JWST} imaging from the COSMOS-WEB and PRIMER surveys, \cite{martorano24} derive qualitatively similar results to this study. They also find that low-mass quiescent galaxies evolve at a similar rate to star-forming galaxies and much more slowly than their high-mass counterparts, although the slopes they obtain at low-stellar masses are somewhat steeper than derived here.


\begin{table}
    \setlength{\tabcolsep}{3pt} %
    \small
    \renewcommand{\arraystretch}{1.0}
    \centering
    \caption{Results from fitting the size-redshift relation to the star-forming and quiescent galaxy samples over the low-mass, high-mass and full-mass range, where $B_{z}$ is the normalisation and $\beta$ is the slope (see Eq. \ref{evol_eq}).}
    \label{tab:size_evol}
    \begin{tabular}{c|cc|cc}
        \hline
        & \multicolumn{2}{c|}{star-forming} & \multicolumn{2}{c}{quiescent} \\
        \midrule
        & $B_{z}$ & $\beta$ & $B_{z}$ & $\beta$ \\
        \hline
        low  & $1.84 \pm 0.03$ & $0.25 \pm 0.03$ & $1.33 \pm 0.02$ & $0.24 \pm 0.08$ \\
        high & $4.07 \pm 0.07$ & $0.45 \pm 0.02$ & $3.07 \pm 0.06$ & $1.14 \pm 0.02$ \\
        all  & $1.89 \pm 0.03$ & $0.24 \pm 0.02$ & $1.60 \pm 0.03$ & $0.42 \pm 0.11$ \\
        \hline
    \end{tabular}
\end{table}

According to our best-fitting relations, the median sizes of low-mass star-forming and quiescent galaxies both increase by only $\simeq 0.08$ dex between $z\simeq 2$ and $z\simeq 0.5$. In contrast, size evolution in the high-mass bin is more rapid, with star-forming galaxies and quiescent galaxies increasing in size by $\simeq 0.14$ dex and $\simeq 0.34$ dex, respectively. The particularly rapid size growth of the high-mass quiescent galaxies is likely to reflect the impact of minor mergers, which are able to increase the half-light radii of galaxies without significantly increasing their stellar mass \citep[e.g.][]{newman2012,ross_size_2013,ownsworth_mergers_2014,buitrago_2017,suess2020,hamadouche22}.

\section{Discussion}\label{discussionc5}

In the previous Section \ref{resultsc5}, we report new size-mass relations for our star-forming and quiescent galaxy samples in PRIMER. Here, we focus on the relationships between stellar mass, size and S\'ersic index, within the context of galaxy quenching mechanisms.

\subsection[The evolution of the galaxy stellar-mass function from \hbox{$\mathit{textbf{z = 0.25}}$} to \hbox{$\mathit{\textbf{z = 2.25}}$}]{The evolution of the galaxy stellar-mass function from \hbox{$\mathit{\textbf{z = 0.25}}$} to \hbox{$\mathit{\textbf{z = 2.25}}$}}

In Section \ref{gsmfc5}, we presented the number densities of quiescent and star-forming galaxies derived from the \textit{JWST} PRIMER survey. These results are in good agreement with the results presented in \cite{mcleod2021_gsmf} and \cite{santini2022}, especially given the difference in survey area (i.e. $\simeq 300$ sq. arcmin versus $\simeq 3$ sq. degrees \& 1000 sq. arcmin, respectively). Within the PRIMER data, we find that the star-forming GSMF can be well described by a single Schechter function that displays only mild evolution in its normalisation out to $z\simeq 2$. This result is well established and our determination of the star-forming GSMF is in good agreement with those of a wide variety of previous studies \citep[e.g.][]{fontana_downsizing, muzzin_2013,davidzon_gsmf}.

In contrast, it can be seen from Fig.~\ref{fig:smfs} that the quiescent GSMF evolves rapidly and is better described by a double Schechter function over the full redshift range of our sample. The two components of the double Schechter function are necessary to reproduce the higher numbers of low-mass quiescent galaxies beginning to enter the population at $z = 2.25$ and likely reflect the two distinct mechanisms by which galaxies quench; environmental quenching, which is independent of mass, and internal quenching, which is dependent on mass \citep{PengQuenching,papovich2018,santini2022}. 

The data presented in this paper extend 1-dex lower in stellar mass than \cite{mcleod2021_gsmf}, and provide the first demonstration with \textit{JWST} of an environmental upturn at $z\geq 2.0$.  Within the PRIMER survey alone, the number densities of quiescent galaxies at higher redshift make it difficult to confirm whether this signature of environmental quenching is also present at $z>2.5$. This open question will require wider-area imaging with sufficient depth to robustly select high-redshift quiescent galaxies down to $\logM\simeq 9$. 

\subsection{Environmental or internal quenching?}\label{envquench?}

Extreme events such as mergers, or interactions involving high-speed fly-bys between galaxies, can affect the morphology of galaxies and increase their S\'ersic indices as they transform from disc-dominated to spheroid-dominated \citep{moore98_gal_harassment}. Disc instabilities in galaxies can also increase S\'ersic index; gravitational instabilities redistribute angular momentum, causing material to be pushed towards the centre and resulting in a bulge-dominated galaxy \citep[e.g.][]{kormendy2004,brennan2015}.

In \cite{sandles2023}, the authors found evidence for a very low-mass quiescent galaxy within an over-density at $z \sim 2$ (with a stellar mass of $\logM \sim 9.0$), as well as two other more massive candidate quiescent galaxies in the nearby region, with properties consistent with environment-driven quenching. Their results demonstrate that a galaxy's environment can still be responsible for quenching at early times, and these observations lend support to our results demonstrating an environment-driven upturn in the quiescent GSMF at $z \sim 2.0$. 
This suggests that the observed distribution in size and S\'ersic index in our PRIMER sample is linked to different evolutionary pathways for low-mass and high-mass quiescent galaxies. 

\cite{cutler2023} study the sizes of quiescent galaxies at cosmic noon ($1 < z< 3$) selected from the PRIMER and UNCOVER surveys. They also find that low-mass and high-mass quiescent galaxies form two distinct populations of galaxies, suggesting that there may be two separate pathways of galaxy formation and quenching at these redshifts, hinted at by differences in axis ratio, S\'ersic index and a constant average size over the low-mass range.
The size-mass relations and the trends observed between various morphological parameters derived in \cite{cutler2023} at cosmic noon are in very good agreement with the results reported in this paper at $0.25 <z < 2.25$. 

Large samples of star-forming, quiescent and post-starburst galaxies show that the number density of low-mass quiescent galaxies increases rapidly towards lower redshift \citep{papovich2018}, with some studies finding a sharp upturn in the number of $\logM<10$ quiescent and PSBs in more dense environments up to $z < 2$ \citep[][]{taylor2023}. 

\subsubsection{Feedback processes}

The high-mass quiescent galaxies in our sample have stellar masses, sizes and S\'ersic indices consistent with quenching via AGN feedback, coupled with growth due to minor mergers. Our results are consistent with the idea that feedback-driven quenching in massive galaxies can result in larger sizes \citep{ubler2014}. \citet{dubois2013} find that, using Adaptive Mesh Refinement code \textsc{Ramses}, AGN feedback can reproduce the observed scaling relations up to $ z\sim 2$. Similarly, \citet{dubois2016} examine the role of AGN feedback on galaxy morphology over a range of redshifts using the hydrodynamical cosmological simulation, \textsc{HorizonAGN}, with and without an AGN component. The authors find that incorporating AGN feedback reproduces the observed size-mass relation of massive quiescent galaxies, albeit with a shallower slope than found by \cite{3dhst_vdw} and broadly consistent with the one found in this work. The lack of star-forming galaxies with comparable S\'ersic indices to the high-mass quiescent galaxies in our sample indicates that internal quenching mechanisms are likely responsible for the high values of S\'ersic index observed for massive quiescent galaxies. 

\subsubsection{Environment driven processes}

At lower redshifts, the progenitors of low-mass quiescent galaxies are proposed to be star-forming galaxies that have been quenched in high-density environments \citep{boselligavazzi2006}. As discussed in \cite{kuchner2017}, galaxies in high-density environments such as clusters are affected by the intra-cluster medium (ICM); as a galaxy falls into a cluster, the pressure of the ICM is greater than the gravitational force between the gas and stellar disk within the galaxy, and gas will be stripped \citep[ram-pressure stripping,][]{gun_gott1972}. This results in the disk of the galaxy fading and reddening over time as the stars within it age, and galaxies falling into cluster centres are expected to have quenched within 1-3 Gyr (at most, 6 Gyr) of entering the dense environment \citep[see e.g.][]{wetzel2013,hirschmann2014,wright2019}. This may explain the lower SFRs and older stellar populations possessed by low-mass quiescent galaxies in high-density environments \citep[see also][]{tomicic_jellyfish,koshy_jellyfish,alberts2023}.
Recent studies combining spatially resolved spectroscopy of so-called ``jellyfish'' galaxies with simulations suggest that star-formation quenching in cluster environments can significantly affect the morphology of an infalling galaxy population on timescales of a few Gyr \citep{marasco_jellyfish}. 

In cluster environments, galaxies close to the centre of the cluster have had more time to lose their star-forming outskirts, resulting in smaller sizes, as observed in the optical, due to the reddening of the disk. However, tidal stripping is also effective at removing some of the loosely bound matter (stars) from the outskirts of galaxies \citep{boselligavazzi2006}, and together these processes can work to quench the galaxy \textit{outside-in}, where the remaining gas towards the centre of the galaxy is used up in star-formation, with the galaxy eventually becoming red and passive but still possessing a (now slightly smaller) disk component, as observed in our PRIMER sample of low-mass quiescent galaxies. 

\cite{schaefer2017} study a sample of 201 star-forming galaxies from the SAMI Galaxy Survey, investigating spatially resolved properties of the environmental quenching of star-formation in galaxies at $0.001 < z < 0.1$. By examining the dust-corrected H$\alpha$ line profiles of these galaxies, and thus the star-formation gradients, the authors find that the fraction of galaxies with centrally-concentrated star formation increases with environment density (for fifth nearest neighbour local environment densities). This result suggests that star formation is suppressed outside-in, such that quenching occurs on the outskirts of the galaxies in dense environments, consistent with mechanisms such as ram-pressure stripping. This scenario is consistent with the galaxies transforming from disc/spirals to S0-type and may explain the large number of red, S0-type galaxies in clusters \citep[e.g.][]{abraham1996,donofrio2015}. 

In the local Universe, the size-mass relation of quiescent galaxies also reflects that observed at higher redshift; towards lower stellar masses, the size-mass relation flattens. Dwarf early-type galaxies appear to have sizes consistent with low-mass star-forming galaxies in the Fornax cluster, and combining these results with the colour-surface brightness relation appears to support a scenario that is consistent with quenching via gas removal \citep{janz_virgo_fornax}, although they find no evidence for an ageing stellar population imprinted on the early-type sample. Studies also find that the size evolution of galaxies in cluster environments is slower in comparison to the field \citep{papovich2012}. 

Additionally, mechanisms that remove the gas from galaxies have been suggested to contribute to the quenching of low-mass galaxies in dense environments. Locally, \cite{samuel2023} find evidence for fast (< 1 Gyr) quenching by ram-pressure stripping of satellite galaxies within dense environments (within $\sim 2$ Mpc from Milky Way-like hosts) in simulations of the Local Group. However, this scenario is highly dependent on distance as ram pressure increases sharply towards the centre of the halo; a galaxy can spend a lot of time ($\sim 1-3$ Gyr) falling into the cluster,  but is then increasingly affected by ram pressure the closer it gets to the cluster centre. 

These trends have also been seen at higher redshifts; up to $z < 1$, results suggest that low-mass quiescent galaxies sit in preferentially higher density environments than their high-mass counterparts, and the number of low-mass star-forming galaxies in dense environments increases between $z \sim 1$ to $z = 0.4$ \citep[e.g.][]{kovac2014,moutard2018}. 
Evidence for fast and slow environmental quenching pathways have been discussed at low redshift, with some studies suggesting that major mergers or ram-pressure stripping of low-mass galaxies in the central regions of the cluster centre can cause quenching within a few hundred Myr \citep[see e.g][]{moutard2016,socolovsky2018}. Internally-driven mechanisms such as star-formation outflows can also interrupt the cold gas supply and quench satellite galaxies before effects such as ram-pressure stripping can take place \citep[overconsumption, see][]{mcgee2014}.

In Fig. \ref{fig:size_redshift_evol}, we show the evolution of sizes for star-forming and quiescent galaxies at $\logM < 10$, $\logM>10$ and over the full stellar-mass range in each of the three panels, respectively. The middle panel confirms previous literature results at high stellar masses; massive star-forming galaxies ($R_{\mathrm{e}} \propto (1 + z)^{-0.45}$) evolve much less dramatically than massive quiescent galaxies ($R_{\mathrm{e}} \propto (1 + z)^{-1.14}$) over the full redshift range. At low-stellar masses, we observe a different picture; both star-forming and quiescent galaxies evolve similarly from $z = 2.25$ to $z = 0.25$, suggesting that the quiescent population is unaffected by mergers or interactions over this time. 

If we consider a low-mass star-forming galaxy with a size of $\sim 1.2$ kpc entering a dense environment at redshift $z < 3$, it will have undergone environment-driven quenching after spending around 1-3 Gyr in this dense environment. This process could plausibly place the galaxy on the observed size-redshift relation (see Fig. \ref{fig:size_redshift_evol}). This scenario is consistent with cluster infall times from simulations and local Universe observations, and does not significantly affect the observed size of the galaxy. Passive evolution (no mergers or interactions) would allow enough cosmic time for the galaxy to begin building up a ``pseudo'' bulge, in agreement with the slightly higher S\'ersic indices of low-mass quiescent galaxies compared to star-forming galaxies found in this work. 

These results suggest that the progenitor population of low-mass quiescent galaxies are low-mass or dwarf star-forming galaxies that quench via environment-driven mechanisms. Further investigation into the local environment of such galaxies, as well as more sophisticated modelling of quenching at lower stellar masses is needed to provide more quantitative evidence for the effect of environmental quenching mechanisms and their relative timescales.

\section{Conclusions}\label{conclusionsc5}

In this work, we explore the relationships between stellar mass, size and morphology at $0.25 < z < 2.25$ for samples of quiescent and star-forming galaxies by utilising high-quality photometric data and imaging from the \textit{JWST} PRIMER survey. The main conclusions can be summarised as follows:

\begin{enumerate}
\setlength\itemsep{0.5em}

\item{We find that the number densities of both quiescent and star-forming galaxies in PRIMER are in good agreement with previous studies of the GSMF over the redshift range $ 0.25 < z < 2.25$. Our results provide new constraints on the quiescent GSMF at low stellar masses at $z\geq 1.5$, providing the first robust demonstration with \textit{JWST} of the upturn at $\logM\leq10$ induced by environmental quenching at $z\simeq 2$.}

\item{We report new size-mass relations for star-forming and quiescent galaxies over the redshift interval $0.25 < z < 2.25$. We confirm that the slope of the size-mass relation for star-forming galaxies does not evolve significantly over cosmic time, although the normalisation does evolve to larger sizes at later times.
At all redshifts, quiescent galaxies display smaller median sizes than star-forming galaxies at fixed stellar mass.}

\item{The slope of the size-mass relation for low-mass quiescent galaxies is indistinguishable from that of the star-forming galaxy relation and shows little sign of evolution within the redshift range studied. Likewise, we find that the redshift evolution in the median size of low-mass quiescent galaxies \hbox{($R_{\mathrm{e}} \propto (1+z)^{-0.24 \pm 0.08}$)} is indistinguishable from that of star-forming galaxies of the same stellar mass.}

\item{In constrast, and in agreement with previous results, we find that the redshift evolution of the median size of high-mass quiescent galaxies is significantly more rapid 
\hbox{($R_{\mathrm{e}} \propto (1+z)^{-1.14 \pm 0.02}$)} than either low-mass quiescent or star-forming galaxies. Moreover, we find that high-mass quiescent galaxies have significantly higher S\'ersic indices than low-mass quiescent galaxies across the full redshift range of our sample. In each redshift bin, low-mass quiescent galaxies have median S\'ersic indices that are intermediate between star-forming galaxies and high-mass quiescent galaxies.}


\item{Overall, our results indicate that low-mass and high-mass quiescent galaxies are distinct populations that have likely been quenched by different processes. The GSMF, sizes and morphologies of low-mass quiescent galaxies strongly suggest that they are quenching via environmental mechanisms, and we show for the first time with \textit{JWST} that this is true out to $z\simeq2$. In contrast, high-mass quiescent galaxies have morphologies and sizes consistent with quenching via internal processes such as AGN feedback, coupled with size growth via dissipationless minor mergers, in agreement with results from previous observations and simulations.}

\end{enumerate}

Using deep near-infrared imaging from the PRIMER survey, we have shown that the quiescent GSMF exhibits the low-mass upturn associated with environmental-driven quenching mechanisms. We have also demonstrated that high-mass quiescent galaxies display morphologies and sizes evolution consistent with the expectations of quenching via internal mechanisms, coupled with growth via minor mergers. In contrast, we find that low-mass quiescent galaxies exhibit size-mass relations and morphologies qualitatively consistent with quenching via infall into cluster environments.

Upcoming, large-scale, near-IR spectroscopic surveys such as MOONRISE \citep{moons} will provide additional information on the kinematics, metallicities and ages of quiescent galaxies, together with improved constraints on their star-formation histories. Moreover, spectroscopic surveys with high completeness will also provide crucial constraints on the local density in which individual galaxies reside, laying the groundwork for a more solid understanding of the dominant quenching mechanisms at cosmic noon.

\section*{Acknowledgements}

M. L. Hamadouche, R. Begley, and C. T. Donnan acknowledge the support of the UK Science and Technology Facilities Council. J.S. Dunlop acknowledges the support of the Royal Society via a Research Professorship. A. C. Carnall acknowledges support from a UKRI Frontier Research Guarantee Grant [grant reference EP/Y037065/1]. F. Cullen and T. M. Stanton acknowledge support from a UKRI Frontier Research Guarantee Grant (PI Cullen; grant reference EP/X021025/1). KEW acknowledges funding from JWST-GO-1837. OA acknowledges the support of STFC grant ST/X006581/1.
The authors would also like to thank S. Manning, S. Wellons \& J. Antwi-Danso for helpful discussions in the writing of this paper.

\section*{Data Availability}
 
All JWST and HST data products are available via the Mikulski
Archive for Space Telescopes (\url{https://mast.stsci.edu}). Additional data products are available from the authors upon reasonable request.



\bibliographystyle{mnras}
\bibliography{main_bib} 




\appendix

\section{Some extra material}

If you want to present additional material which would interrupt the flow of the main paper,
it can be placed in an Appendix which appears after the list of references.


\bsp	
\label{lastpage}
\end{document}